\documentclass[prl,preprint,showpacs,superscriptaddress]{revtex4}
\usepackage{graphicx}
\usepackage{indentfirst}
\usepackage{psfrag}
\usepackage{epsfig}
\usepackage{subfigure}
\usepackage{amsmath}
\usepackage{amssymb}

\def\be{\begin{eqnarray}}
\def\ee{\end{eqnarray}}
\def\r{{\bf r}}

\def\k{{\bf k}}

\begin{document}
\title{Controlling absorption resonances from sub-wavelength cylinder arrays}

\author{Marine Laroche}

\address{Departamento de F\'{\i}sica de la Materia Condensada
and Instituto ``Nicol\'{a}s Cabrera'', Universidad Aut\'{o}noma de
Madrid, E-28049 Madrid, Spain.}\homepage{http://www.uam.es/mole}


\author{Raquel G\'{o}mez-Medina}

\address{Nanophotonics and Metrology Laboratory, \\ Swiss
Federal Institute of Technology, 1015 Lausanne, Switzerland}


\author{Juan Jos\'e S\'aenz}

\address{Departamento de F\'{\i}sica de la Materia Condensada
and Instituto ``Nicol\'{a}s Cabrera'', Universidad Aut\'{o}noma de
Madrid, E-28049 Madrid, Spain.}\homepage{http://www.uam.es/mole}


\begin{abstract}
The absorption and extinction spectra of sub-wavelength cylinder
arrays are shown to present two different kind of resonances. Close
to the Rayleigh anomalies, the diffractive coupling with the lattice
periodicity leads to sharp peaks in the extinction spectra with
characteristic Fano line shapes for both $s$ and $p$-polarizations.
When the material exhibits an absorption line or in the presence of
localized surface plasmon/polaritons, the system is shown to present
resonant absorption with wider and symmetric line shapes. For
$s$-polarization our analysis predict a theoretical limit of 50\% of
absorption. Interestingly, for $p$-polarized light and an
appropriate choice of parameters, a subwavelength cylinder array can
present perfect (100\%) absorption.
\end{abstract}

\date{24, April 2006}
\pacs{42.25.Bs,42.79.Dj,44.40.+a,73.20.Mf}

\maketitle

The study of the extinction spectra of nanoparticles have drawn much
attention recently for their potential applications to chemical and
biological sensors as well as for Surface Enhanced Raman Scattering
(SERS) \cite{review_chem,OJF1,nanoeng}. While for a single
nanoparticle, the excitation of localized surface plasmon/polariton
resonances (LSPR) lead to well defined extinction line shapes at
specific wavelengths\cite{OJF1}, for an assemble of nanoparticles
the spectra can be substantially modified by multiple scattering
effects. For absorbing materials, resonant {\em radiative}
diffraction as well as resonant {\em absorption} processes could
both  lead to sharp peaks in the extinction spectra. Our main
purpose is to understand the relative role of both processes in the
spectra of nanoparticle arrays. This is an important issue from a
fundamental point of view but it is also specially relevant for a
wide range of applications: By Kirchhoff's law \cite{nieto}, the
control of absorption by nanostructured materials is equivalent to
tailor the thermal emission. This is of interest for
thermophotovoltaic applications and for the design of efficient
infrared sources. The recent development of coherent thermal sources
\cite{nature_jj,ol_marine} has stimulated further  work on thermal
emission from  photonic crystals \cite{lin_prb,enoch_apl,prl_marine}
 or nanoparticle arrays
\cite{yannopapas}.

In the visible range, plasmon resonances on linear array of metallic
particles have been intensively studied considering the influence of
different parameters on the extinction spectra
\cite{krenn_prl,zhang_durant,schatz}.
Enhanced absorption by dielectric particles resulting of
phonon-polariton resonances has also been demonstrated in the
infrared \cite{apl_sic}. Peaks in the extinction spectra are
commonly associated to the excitation of LSPR modified by the
diffractive coupling with the lattice periodicity
\cite{schatz,apl_sic}. For cylinder arrays, $s$-polarized
electromagnetic radiation (with the electric field parallel to the
cylinder axis) can not excite any surface plasmon. However, the
extinction spectra of subwavelength cylinder arrays of non-absorbing
dielectric materials present sharp peaks even for $s$-polarization
\cite{raquel}.

In this letter, we study the extinction spectra of sub-wavelength
cylinder arrays. As we will see, the spectra present two different
kind of resonances. Close to the Rayleigh anomalies, the diffractive
coupling with the lattice periodicity leads to sharp peaks in the
extinction spectra with characteristic Fano line shapes for both $s$
and $p$-polarizations. These {\em geometric} resonances, associated
to radiative coupling in absence of absorption, may lead to resonant
absorption for appropriate grating parameters. We analytically
derive the conditions for resonant absorption as a function of the
geometry and material's parameters. Another kind of absorption
resonances, with wider and symmetric line shapes,  appears when the
material exhibits an absorption line or in the presence of LSPRs. In
contrast with geometric resonances, these absorption peaks are
almost isotropic with a weak dependence on the angle of incidence
and geometry. We will demonstrate that for $s$-polarization there is
a theoretical limit of 50\% of absorption. Interestingly, we will
show that, for $p$-polarized light and an appropriate choice of
parameters, a subwavelength cylinder array can present perfect
(100\%) absorption.




\begin{figure}[htbp]
\begin{center}
\includegraphics[width=7cm]{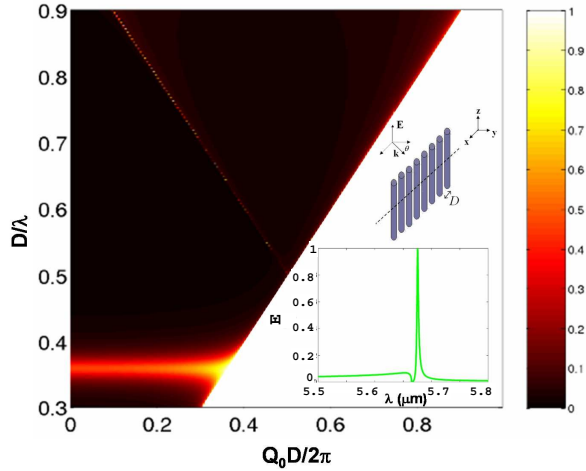}
\caption{({\bf $s$-polarization }) Extinction in s-polarization in a
map of frequency ($D/\lambda$) versus the transversal momentum of
the incoming radiation ($Q_0 = 2 \pi  \sin{\theta} /\lambda$), for
an array of SiC nanocylinders with period $D = 4.5 \mu m$ and radius
$a=0.2 \mu m$. Around $D/\lambda$ =0.36 ($\lambda = 12.5 \mu m$),
there is an isotropic extinction peak due to the absorption line of
SiC. The inset shows the extinction spectrum, which exhibits a
typical Fano line shape, for an incident angle $\theta=15^o$ around
the geometric resonance (close to the first Rayleigh frequency, i.e.
$D/\lambda=1-Q_0D/2\pi$).} \label{res_geos}
\end{center}
\end{figure}

Let us consider an infinite set of parallel cylinders with their
axis along the z-axis (see inset in Fig. \ref{res_geos}), relative
dielectric constant $\epsilon = \epsilon' + i \epsilon''$ and radius
$a$ much smaller than the wavelength. The cylinders are located at
$\r_n = nD {\bf u}_x = x_n {\bf u}_x$ (with $n$ an integer number).
For simplicity, we will assume incoming plane waves with wave vector
$\k_0 \bot {\bf u}_z$ ( i.e. the fields do not depend on the
$z$-coordinate), with $k = \omega/c$ and $\k_0 = k \sin \theta \
{\bf u}_x + k \cos \theta \ {\bf u}_y \equiv Q_0 {\bf u}_x + q_0
{\bf u}_y .$ The reflectance, $R$ and transmittance, $T$ of the
cylinder array can be calculated by using a standard multiple
scattering approach in the dipolar approximation \cite{Foldy}, as
described in Ref. \cite{raquel}.
%
We define the absorptivity, $A$, as: $ A \equiv 1 - R - T $ and the
normalized extinction, $E$, as the ratio between
 scattered plus absorbed powers and incoming power
(notice that, below the onset of the first diffraction beam, the
extinction is simply  given by the sum of absorption and specular
reflection, $E=A+R$ ).

Let us first consider the simpler case of  $s$-polarized electromagnetic waves.
For a single sub-wavelength cylinder, the polarizability is given by
\be
 \frac{1}{k^2\alpha_{zz}} = \left\{
 C (\epsilon'-1) + \cdots\right\} - i\left\{\frac{1}{4} +
 C \epsilon'' + \cdots\right\} \label{unoa}
 \ee
 where $C^{-1} \equiv \pi (ka)^2|\epsilon-1|^2$.
It is worth noticing that $\left(-\Im\left\{1/(k^2\alpha_{zz})
\right\}-\frac{1}{4}\right) \propto \epsilon''$, so,  in absence of
absorption ($\epsilon''=0$), the expression above is
 consistent with the optical theorem \cite{raquel,raquel2,comment_schatz}.
Multiple scattering effects, due to the presence of the
 other scatterers, can be included in a
renormalized polarizability, $\widehat{\alpha}_{zz}$
\cite{raquel,raquel2}. Generalizing the results of Ref.
\cite{raquel} to include absorption we find
 \be
\left(k^2 \widehat{\alpha}_{zz}\right)^{-1}
&=& \Re\left\{ \frac{1}{ k^2
\alpha_{zz}} - G_b \right\} -i \left\{ C\epsilon'' + \Im\{G\}
\right\}
 \ee
where $G_b$ is the depolarization term  \cite{raquel,raquel2,meier},
defined as
 $G_{b} = \lim_{\r \rightarrow \r_0}
\left[G(\r)-G_0(\r,\r_0)\right]$,  being $G$ and $G_0$  the Green
function of the periodic array and the free-space respectively, and
$\Im\{G\}\equiv \Im\{G(0)\}$.
The absorptivity and normalized extinction can now be written in
terms of $\widehat{\alpha}_{zz}$ as: \be
A^{(s)} &=& \frac{k^4|\widehat{\alpha}_{zz}|^2}{Dq_0} C \epsilon'' \\
E^{(s)} &=& A^{(s)} + \frac{k^4|\widehat{\alpha}_{zz}|^2}{Dq_0}
\left(\Im\{G\}
-\frac{1}{4Dq_0} \right). \ee 

\begin{figure}[htbp]
\begin{center}
\psfrag{D}{\hspace{-0.5cm} \small{$\mathbf{\lambda \: \rm (\mu
m)}$}} \psfrag{X}{\hspace{-0.6cm}
\footnotesize{$\Im\{\alpha/a^2\}$}} \psfrag{Y}{\hspace{-0.9cm}
\small{$\Re\{a^2/\alpha\}$}} \psfrag{A}{\hspace{-1.1cm}
\footnotesize{absorption factor}} \psfrag{B}{\hspace{-1.1cm}
\footnotesize{extinction factor}} \psfrag{F}{\hspace{-0.2cm}
\small{$\lambda_P$}} \psfrag{E}{\hspace{-0.2cm} \small{$\lambda_T$}}
\psfrag{G}{\hspace{-0.1cm} \small{$\lambda \: \rm (\mu m)$}}
\includegraphics[width=8cm]{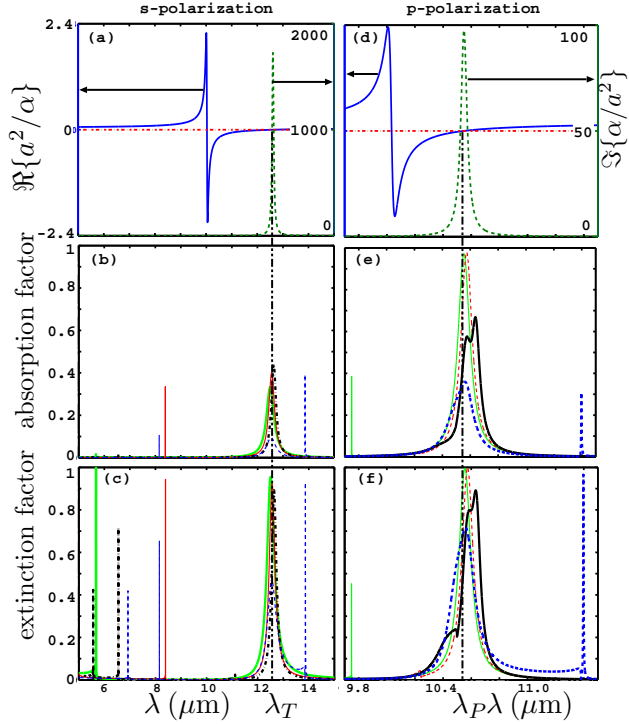}
 \caption{({\bf $s$-polarization }) (a): Real part of $1/\alpha_{zz}$ (-) and
imaginary part of $\alpha_{zz}$ (- -) versus $\lambda$ for SiC
cylinders. Absorption resonance is due to a zero ($-\cdot-$) of
$\Re\{1/\alpha_{zz}\}$, which coincides with a maximum of
$\Im\{\alpha_{zz}\}$. (b): Absorption spectrum versus $\lambda$ for
arrays of cylinder of radius $a = 0.2 \: \mu m$, for an incident
angle $\theta =15^o$ and for several periods \: $D=4.5 \: \mu m$
(green bold solid line), $D=6.67 \: \mu m$ (red solid line), $D=8.83
\: \mu m$ (black dotted-dash line), $D=11 \: \mu m$ (blue dashed
line). (c): Extinction spectrum for the same parameters than (b).
\\ ({\bf $p$-polarization }) (d): Real part of $1/\alpha_{p}$ (-) and imaginary part of
$\alpha_{p}$ (- -) versus $\lambda$ for SiC cylinders. (e):
Absorption spectrum versus $\lambda$ for arrays of cylinders of
radius $a=0.5 \: \mu m$ and different periods \: $D= 5.71 \: \mu m$
(green solid line), $D=6.0 \: \mu m$ (red dashed line), $D=6.15 \:
\mu m$ (black bold solid line), $D=6.667 \: \mu m$ (blue bold dashed
line), (f): Extinction spectrum for the same parameters than (e).}
\label{alpha_abs_ext_sp}
\end{center}
\end{figure}

In order to illustrate the main physics involved in the different
resonant phenomena, we  will  consider a  typical dielectric
constant given by: $ \displaystyle{ \epsilon=\epsilon_{\infty}
\left(\omega_L^2-\omega^2-i \gamma \omega \right)
\left\{\omega_T^2-\omega^2-i \gamma \omega \right\}^{-1}.}$ This is
a standard form for a polar material (Lorentz model), for a metal,
the Drude model is recovered taking $\omega_T=0$ and $\omega_L =
\omega_p$. Silicon carbide (SiC) nanowires  provide a simple model
system: its dielectric constant is given by this form with the
following parameters \cite{palik}: $\epsilon_{\infty}=6.7$,
$\omega_L=1.825 \: 10^{14} \: \rm rad.s^{-1}$, $\omega_T=1.494 \:
10^{14} \: \rm rad.s^{-1}$, $\gamma=8.9662 \: 10^{11} \: \rm
rad.s^{-1}$.
Figure \ref{res_geos} displays the extinction in the infrared range,
in a map of frequency ($D/\lambda$) versus the transversal momentum
of the incoming radiation ($Q_0$), for an array of SiC cylinders
(with $D=4.5 \: \mu m$ and $a=0.2 \: \mu m$). The extinction spectra
show two different kinds of resonances: {\em i)} close and below the
Rayleigh frequency $\omega \rightarrow \omega_1^{(-)} =
c|Q_0-2\pi/D|$ the spectra present a very sharp peak with a strong
dependence on the angle of incidence, and {\em ii)} a broad peak
close to the absorption line of SiC ($\omega \approx \omega_T$)
which is almost isotropic. The peaks in the extinction spectra
correlate with corresponding maxima in the absorption. However, the
relative strength of the peaks depends on the material and
geometrical parameters. This is illustrated in
Fig.\ref{alpha_abs_ext_sp} where we have plotted the real part of
$1/\alpha_{zz}$  and the imaginary parts of $\alpha_{zz}$ (2a), the
absorptivity (2b) and the extinction (2c) versus the wavelength for
an incident angle $\theta=15^o$ an for different lattice constants,
$D$ (the inset in Fig. 1 corresponds to a zoom of the bold solid
line in Fig. 2c).

The different extinction/absorption resonances resemble the recently
discussed ``lattice'' \cite{raquel,JdA1} and ``site'' \cite{JdA2}
resonances in absence of absorption. Resonant processes arise
when the real part of $1/(k^2\alpha_{zz})- G_b$ vanishes.
Approaching the threshold of the first propagating (diffraction)
channel, the real part of $G_b$ goes to infinity as $ \approx
(\omega_1^2-\omega^2)^{-1/2}c/(2D)$ and can compensate exactly the
real part of $1/k^2\alpha_{zz}$ giving rise to a ``geometric''
resonance \cite{raquel,raquel2}. This can only happen when the real
part of the polarizability is positive ($\epsilon' > 1$). As a
consequence, in $s$-polarization there is no resonant absorption for
metallic cylinders. The typical Fano line shape of these resonances
is illustrated in the inset of Fig. \ref{res_geos}.  While it is
possible to obtain a 100\% extinction of the beam ( due to the
reflection resonances which appear when the absorption is weak
\cite{raquel} ), the maximum absorption $A_{max}$ is limited to a 50
\% (see Fig.\ref{alpha_abs_ext_sp}(b,c)). It is easy to show that
the highest absorption, $A_{max}= 1/2$, takes place below the first
Rayleigh frequency when \be \displaystyle{ C \epsilon''=
\frac{\epsilon''}{\pi (ka)^2|\epsilon-1|^2} = \frac{1}{2Dq_0}.}
\label{Important1} \ee This is one of the central results of this
work. We can also notice that at the Rayleigh frequency, $\omega =
\omega_{1}$, the array of cylinders is transparent as the absorption
and the extinction goes to zero.

Material or ``site'' resonances are associated to the zeros of the
real part of $1/\alpha_{zz}$. As shown in
Fig.\ref{alpha_abs_ext_sp}(a), it exhibits two zeros, one close to
$\omega \simeq \omega_T$ the other to $\omega \simeq \omega_S =
(\epsilon_{\infty}\omega_L^2-\omega_T^2)/(\epsilon_\infty-1)$. The
second one corresponds to a small value of $\Im\{\alpha_{zz}\}$,
which leads to a weak absorption. The broad resonances in Figure
\ref{alpha_abs_ext_sp}(b) and (c) coincide with a maximum of
$\Im\{\alpha_{zz}\}$ and correspond to a resonant absorption of the
material. Then, they should  not be very sensitive to the lattice
parameters and order. However, the maximal absorptivity value of
$1/2$ can only be reached at the condition given by eq.
\ref{Important1}. Close to the Rayleigh frequencies, there is a
small blue or red-shifted depending of the value of the real part of
$G_b$ (see Fig 2a. of ref \cite{raquel}).

Let us now consider an incoming wave with the magnetic field
parallel to the cylinder axis ($p$-polarized wave).
Following the notation of ref. \cite{raquel} we can now define the
renormalized polarizabilities of the effective dipoles pointing
along the x and y axis: $ \widehat{\alpha}_{xx} =
\left(1+\alpha_{xx}
\partial^2_y G_b \right)^{-1} \alpha_{xx} $, $ \widehat{\alpha}_{yy}
= \left(1+\alpha_{yy} \partial^2_x G_b \right)^{-1} \alpha_{yy}$,
being $\partial^2_{x,y} G_b$  the depolarization terms due to the
components x and y of the electric field scattered by all the
dipoles (except the considered one)
 and $\alpha_{xx}=\alpha_{yy}=\alpha_p$
 the polarizability of a cylinder in p-polarization,
\be
 \frac{1}{k^2\alpha_{p}} = \left\{
 \frac{C}{2} (|\epsilon|^2-1) + \cdots\right\} - i\left\{\frac{1}{8} +
 C \epsilon'' + \cdots\right\}.
 \ee
 The general
expression for the absorption is:
\be A^{(p)}\,=\, \frac{k^2}{Dq_0} \left(
Q_0^2|\widehat{\alpha}_{yy}|^2+q_0^2|\widehat{\alpha}_{xx}|^2
\right) C \epsilon''  .\ee
Hence, in p-polarization, the absorption results from the sum of the
contributions of two dipoles, one pointing in the x-direction (with an
effective polarizability $\widehat{\alpha}_{xx}$) and one pointing in the
y-direction (with an effective polarizability $\widehat{\alpha}_{yy}$).

At the threshold of the first propagating order, there is a resonant
coupling of electric dipoles pointing along the $y$-axis which leads
to the divergence of $\Re\{\partial_x^2 G_b\} \approx -
(\omega_1^2-\omega^2)^{-1/2}\omega_1^2/(2cD)$ at the Rayleigh
frequencies (in contrast $\Re\{\partial_y^2 G_b\}$ remains finite).
Close to the first Rayleigh anomaly ($\omega \lesssim \omega_1$) and
providing that $\Re\{1/\alpha_p>0\}$ \cite{raquel}, the absorption
contains two parts: a resonant part due to the resonance of
$\widehat{\alpha}_{yy}$ (and similar to the case in s-polarization),
and the term due to $\widehat{\alpha}_{xx}$ to which we will refer
as a "background contribution".
 For p-polarization, the resonant part has also a maximal value of $1/2$ and reaches this maximum
when  \be \displaystyle{ C \epsilon''=
\frac{Q_0^2}{k^2}\frac{1}{2Dq_0}.} \label{Important2} \ee Depending
on the contribution of the non-resonant part, the absorption can
reach a value higher than $1/2$.

Site resonances for $p$-polarization are again associated to zeros
of the real part of $1/\alpha_{xx}=1/\alpha_{yy}=1/\alpha_p$.
$\Re\{1/\alpha_p\}$ exhibits two zeros (see Fig.
\ref{alpha_abs_ext_sp} (d)): one close to $\omega \simeq \omega_S$,
but corresponding to a small value of $\Im\{\alpha_p\}$ and to
$\epsilon'=+1$ and thus a weak absorption (for SiC, $\lambda_S =
10.0 \mu m$), the other close to $\omega \simeq \omega_P$
($\epsilon' \approx -1$) which gives rise to ( phonon-polaritons or
plasmon-polaritons for metals) LSPR  (for SiC, $\lambda_P = 10.57
\mu m $). In contrast with the geometric resonances, the LSPR  will
lead to the resonance of both dipoles. As each dipole can contribute
up to 1/2, the absorption may reach $100\%$.

\begin{figure}[htbp]
\begin{center}
\psfrag{A}{\hspace{-0.2cm}{$Q_0 D/2\pi$}}
\psfrag{B}{\hspace{-0.8cm}{$D/\lambda$}}
\includegraphics[height=8cm]{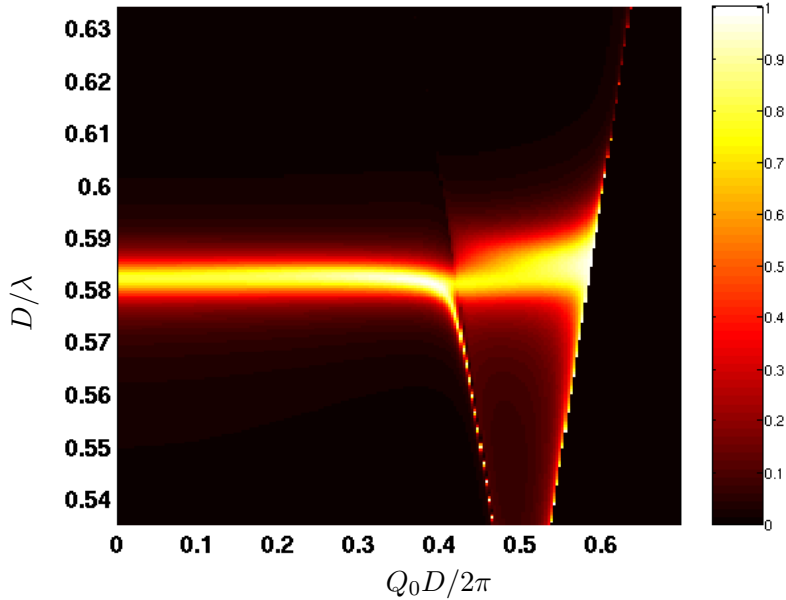}
\caption{({\bf $p-$polarization}) : Extinction  map for an array of
SiC cylinders with a period $D=6.15 \: \mu m$, a radius of $a=0.5 \:
\mu m$. The localized surface phonon-polaritons resonance,  at the
wavelength $\lambda \simeq \lambda_P$, is isotropic.}
\label{res_partp}
\end{center}
\end{figure}

We have to notice that unless $\Re\{\partial_x^2
G_b\}=\Re\{\partial_y^2 G_b\}$, the two effective dipoles will not
resonate exactly at the same frequency, thus this value of the
absorption maximum will be very sensible to the lattice parameters.
Together with this condition , it can be easily shown that the other
condition to reach the maximal absorption is $ q_0 = Q_0 (\theta
=45^o)$. We represent in Fig.\ref{alpha_abs_ext_sp}(e) and (f),
respectively the absorption and extinction spectra of an array of
SiC cylinders with a radius of $a=0.5 \: \mu m$ and for an incident
angle of $\theta = 45^o$ and several periods. Similarly to
s-polarization, there are two kinds of resonances, (1) geometric
resonance characterized by very sharp peaks at wavelengths close to
Rayleigh frequencies, (2) a broader {\em double peak} at a
wavelength $\lambda \simeq \lambda_P$. This double peak corresponds
to the resonances of the $y-$ and $x-$dipoles, (red or blue) shifted
depending on the sign (+ or -) of $\partial_x^2 G_b$ and
$\partial_y^2 G_b$. As discussed above, $\approx$ 100\% absorption
takes place only when $\partial_x^2 G_b \approx \partial_x^2 G_b$.
It is worth noticing that in the wavelength range where
$\Re\{1/\alpha_p<0\}$ (i.e. $|\epsilon|<1$), there are no geometric
resonances. This explains the anomalous shape of the LSPR extinction
peak for $D=6.15 \: \mu m$ (black bold solid line in
Fig.\ref{alpha_abs_ext_sp}(f)), as the Rayleigh frequency ($\lambda
=10.5 \: \mu m$) lies in a range where $\Re\{1/\alpha_ d\}\lesssim
0$. At this wavelength, we can just observe the ($y-$dipole)
Rayleigh transparency dip in the extinction spectrum (see also Fig.
\ref{res_partp}). These results provide a simple and analytical
explanation of the anomalous shape of the extinction peaks observed
in recent numerical simulations \cite{schatz}.

In conclusion, this paper gives a simple analytical method to derive
the optical properties of sub-wavelength cylinders arrays. We have
shown that in s-polarization, absorption resonances can absorb up to
 half of the incident power.  The absorption and extinction spectra for
p-polarization presents a more complex structure due to the
contribution of two orthogonal dipoles. It is remarkable that, as we
have shown, by an appropriate choice of parameters a subwavelength
cylinder array can become a perfect absorber. We believe that our
analysis paves a new way for the nanoengineering of chemical and
biological sensors and photo-thermal devices based on nanoparticle
arrays.

%
We thank  S. Albaladejo, J. Garc\'{\i}a de Abajo and O. J. F. Martin
for interesting discussions. This work has been supported by the
Spanish MEC (Ref. No. EX2005-1181), the EU Integrated Project
``Molecular Imaging'' (EU contract LSHG-CT-2003-503259) and the EU
network of excellence "Plasmo-nano-device" (FP6-2002-IST-1-507879).

\end{document}